\documentclass[conference]{IEEEtran}
\usepackage{cite}
\usepackage{amsmath,amssymb,amsfonts}
\usepackage{algorithmic}
\usepackage{graphicx}
\usepackage{textcomp}
\usepackage{xcolor}
\def\BibTeX{{\rm B\kern-.05em{\sc i\kern-.025em b}\kern-.08em
    T\kern-.1667em\lower.7ex\hbox{E}\kern-.125emX}}
\begin{document}

\title{Assisted Text Annotation Using Active Learning to Achieve High Quality with Little Effort}

\author{\IEEEauthorblockN{Franziska Weeber\IEEEauthorrefmark{1}, Felix Hamborg\IEEEauthorrefmark{2}, Karsten Donnay\IEEEauthorrefmark{3}\IEEEauthorrefmark{2} and Bela Gipp\IEEEauthorrefmark{4}}
\IEEEauthorblockA{\IEEEauthorrefmark{1}University of Konstanz, Germany, \IEEEauthorrefmark{2}Heidelberg Academy of Sciences and Humanities, Germany, \\}
\IEEEauthorblockA{\IEEEauthorrefmark{3}University of Zurich, Switzerland, \IEEEauthorrefmark{4}University of Wuppertal, Germany}
franziska.weeber@uni-konstanz.de
}

\maketitle

\begin{abstract}
Large amounts of annotated data have become more important than ever, especially since the rise of deep learning techniques. However, manual annotations are costly. We propose a tool that enables researchers to create large, high-quality, annotated datasets with only a few manual annotations, thus strongly reducing annotation cost and effort. For this purpose, we combine an active learning (AL) approach with a pre-trained language model to semi-automatically identify annotation categories in the given text documents. To highlight our research direction's potential, we evaluate the approach on the task of identifying frames in news articles. Our preliminary results show that employing AL strongly reduces the number of annotations for correct classification of even these complex and subtle frames. On the framing dataset, the AL approach needs only 16.3\% of the annotations to reach the same performance as a model trained on the full dataset. 
\end{abstract}


\section{Introduction and Related Work}

With the rise of machine learning and recently deep learning, annotated data have become more important than ever before. While traditional machine learning relies mostly on handcrafted rules and features, recent approaches use large amounts of annotated data. Not only computer scientists rely on these extensive labeled corpora, they are also needed in other disciplines, e.g., for the large-scale analysis of media bias in political science, genome diagnostics in the medical and biological sciences, and other practical applications such as online customer support bots. 
Large amounts of unlabeled data are available for most tasks, but creating the annotations needed to use them as training sets is costly \cite{settles_2009}. A solution is active learning (AL), where a classifier is trained over multiple iterations by querying batches of only those instances for manual annotation from a large unlabeled pool that are considered most informative for the model \cite{settles_2009}. AL can substantially reduce the number of labeled observations needed to classify the full pool, especially when classes are unbalanced because the minority class will be overrepresented in the queried training data \cite{tomanek_2007, settles_2009}. 

The AL process is structured into two phases: seed set selection and iterative query selection plus training. During \textit{seed set selection}, the initial training set is chosen, e.g., randomly, manually, or based on other criteria such as seed words or cluster membership \cite{tomanek_2007}. Humans label these instances, and then the model is trained on them. In \textit{query selection}, each iteration entails: predicting the class probabilities for the data pool, querying new instances for labeling, and retraining the model. The first task aims to determine which instances to query for labeling next using a measure of informativeness, e.g., the uncertainty of the prediction. Afterward, the classifier is retrained on the updated labeled set until a performance threshold is reached or a predefined number of texts have been annotated \cite{tomanek_2007}. 

Previous work on semi-automated data annotation employing AL suffers from the following shortcomings: (1) superficial or poor results and (2) lack of easy-to-use annotation interfaces. First, prior methods using AL often use outdated methods, such as bag of word representations and SVMs \cite{tomanek_2007, skeppstedt_2016}, not reaching the performance of language models such as BERT or its successors. Second, most approaches do not provide an interface for the annotation \cite{skeppstedt_2016} or still require users to have programming experience \cite{prodigy}. Despite these shortcomings, the usefulness of AL annotation tools for segmentation problems, such as NER or POS tagging, where selecting the correct label is straightforward for an expert, has already been shown \cite{tomanek_2007, skeppstedt_2016}. 

In sum, many approaches have proven to work in theory but are not available to researchers who wish to create a labeled dataset based on their predefined categories. To our knowledge, no tool employs AL and state-of-the-art language models for researchers without detailed knowledge about programming and NLP concepts, and that facilitates the annotation of non-trivial categories, such as media frames.

We propose a tool for assisted, semi-automated data annotation based on state-of-the-art language models and AL. Our approach seeks to (1) reduce the cost needed to create a labeled corpus while retaining (2) high quality of the annotations. Both factors are essential to scientists who require training data, e.g., to train a neural model, and researchers who need labeled corpora for their analyses, e.g., social scientists performing a large-scale content analysis.

\section{Active Learning for Frame Annotation}

Our goal is to devise an annotation tool that enables researchers to create a large, labeled dataset of high quality while strongly reducing annotation cost and effort. The proof of concept in this poster employs our AL framework to demonstrate the effectiveness and (cost-)efficiency in a difficult real-world annotation task comprising frames in news articles. Framing refers to the practice of reporting about topics from a certain perspective to shape readers' opinions. 

Our tool aims to combine the strength of natural language understanding (NLU) achieved by transformer-based language models and an AL approach's efficiency. By keeping the domain expert in the loop, we enable high-quality, semi-automated labeling. We use a simple classifier consisting of a language model with a fully connected layer added on top of the classification token. This architecture demonstrates high capabilities to identify nuanced and small classes. Since we aim to classify entire articles, we use Longformer \cite{beltagy_2020} to retrieve one embedding for each document. Using a pre-trained model, such as Longformer, rather than training a model from scratch is another means we employ to drastically reduce the size of the training data required.

We randomly select our seed set. In the query selection steps, we query data based on how uncertain the model is about its prediction, i.e., with the smallest margin between the two largest logits returned by the model, and retrieve their label. This query strategy is computationally simple but efficient \cite{settles_2009}.

\section{Preliminary Results}

Our evaluation investigates how AL can reduce the amount of data required for training to still achieve high classification performance. 
Conceptually, our method can be applied to any classification task. For this poster, we conduct our preliminary evaluation on the previously mentioned complex labeling task. Specifically, we use the Media Frames Corpus (MFC) \cite{card_2015}, which consists of news articles reporting on five topics published by US outlets and includes labels for 15 generic frames on the sentence and the article level. Labeling frames is difficult, even for human annotators, due to the frames' subtlety and similarity to another. We use the topic \textit{death penalty} and select all 2,458 articles with one of three most frequently used primary frames: \textit{Political} (10\%), \textit{Crime and Punishment} (29\%), and \textit{Legality, Constitutionality, and Jurisdiction} (61\%) . 
We split the dataset in an AL pool (80\%) and evaluate our approach on the test set of the remaining 20\% of all articles. 

To see if AL reduces the number of annotations required, we start by randomly sampling a seed set of 160 documents and then query 40 instances each in ten rounds based on the smallest margin of the prediction. 
As a baseline, we compare the active learner to a passive learner, which selects new instances randomly.
\begin{figure}[htbp]
\centerline{\includegraphics[width=0.4\textwidth]{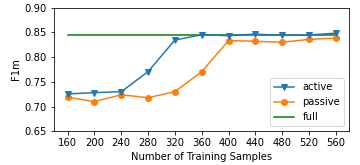}}
\caption{Performance of Active and Passive Learner}
\label{fig1}
\end{figure}
The model trained on the full pool achieves a classification performance of $F1_m=84.6$. Our AL model achieves $F1_m=83.5$ after having labeled 320 examples, which equals four rounds. In comparison, the maximum performance of the passive learner is $F1m=83.4$ with 400 training samples and six rounds, respectively (see fig. 1). On the majority class, the performance is equally good for both learners, but the AL model learns faster for the minority classes. The results highlight that the model can reach high performances trained only on a fraction of the full data pool.

Our future work includes the following expansion of our approach. First, we plan to test the model in a multi-label task on the sentence- or sequence-level, where a phrase has zero or more classes, e.g., frames in MFC. Second, we plan to investigate further strategies for selecting the seed set and the queried sets and test further selection measures, e.g., to increase the diversity and representativeness of the sets to be labeled and thus accelerate the performance increase even more. Last and most important, we need to devise a user interface for non-expert users, e.g., that can be used by researchers even without detailed knowledge on the methodology used for the AL and classification part. 

\section{Conclusion}
We proposed an approach for semi-automated data annotation that uses active learning and a recent language model. Our preliminary evaluation of the approach on a difficult real-world use case showed a reduction of $83.7\%$ in the number of training samples required in comparison to the full data pool and a reduction of $20.0\%$ in comparison to a passive learner while still maintaining high-quality classification performance ($F1_m=83.5$ compared to $84.6$ achieved by the model on all examples). 
We envision integrating this approach into an annotation tool, not requiring knowledge of the methods we used. Both researchers who need data to train their models \cite{hamborg2021} and researchers who wish to perform content analyses could benefit from the tool's cost-efficiency.


\begin{thebibliography}{00}
\bibitem{beltagy_2020} I. Beltagy, M. E. Peters, and A. Cohan, “Longformer: The Long-Document Transformer,” arXiv:2004.05150 [cs], Apr. 2020.
\bibitem{card_2015} D. Card, A. E. Boydstun, J. H. Gross, P. Resnik, and N. A. Smith, “The Media Frames Corpus: Annotations of Frames Across Issues,” in Proceedings of the 53rd Annual Meeting of the Association for Computational Linguistics and the 7th International Joint Conference on Natural Language Processing (Volume 2: Short Papers), Beijing, China, Jul. 2015, pp. 438–444, doi: 10.3115/v1/P15-2072.
\bibitem{prodigy} Prodigy. Explosion AI, 2021. https://prodi.gy/
\bibitem{settles_2009} B. Settles, “Active Learning Literature Survey,” University of Wisconsin-Madison Department of Computer Sciences, Computer Sciences Technical Report, 2009. 
\bibitem{skeppstedt_2016} M. Skeppstedt, C. Paradis, and A. Kerren, “PAL, a tool for Pre-annotation and Active Learning,” Journal for Language Technology and Computational Linguistics, vol. 31, no. 1, pp. 81–100, 2016.
\bibitem{tomanek_2007} K. Tomanek, J. Wermter, and U. Hahn, “Efficient annotation with the Jena ANnotation Environment (JANE),” in Proceedings of the Linguistic Annotation Workshop on - LAW ’07, Prague, Czech Republic, 2007, pp. 9–16, doi: 10.3115/1642059.1642061.
\bibitem{hamborg2021} F. Hamborg and K. Donnay, “NewsMTSC: A Dataset for (Multi-)Target-dependent Sentiment Classification in Political News Articles,” in Proceedings of the 16th Conference of the European Chapter of the Association for Computational Linguistics: Main Volume, Online, Apr. 2021, pp. 1663–1675.

\end{thebibliography}
\end{document}